\pdfoutput=1
\documentclass[11pt,letterpaper]{article}
\usepackage[margin=1in,headheight=36pt,headsep=14pt]{geometry}
\usepackage[T1]{fontenc}
\usepackage{times}
\usepackage{mathptmx}
\usepackage[protrusion=true,expansion=false]{microtype}
\usepackage{amsmath}
\usepackage{booktabs}
\usepackage{graphicx}
\usepackage[table]{xcolor}
\usepackage{tikz}
\usetikzlibrary{positioning,arrows.meta,fit}
\usepackage[font=small,labelfont=bf,skip=4pt]{caption}
\usepackage{enumitem}
\setlist{nosep,leftmargin=*}
\usepackage[numbers,sort&compress]{natbib}
\definecolor{citeblue}{HTML}{3A6EA5}
\usepackage[colorlinks=true,citecolor=citeblue,linkcolor=citeblue,urlcolor=citeblue]{hyperref}
\usepackage[capitalise,nameinlink]{cleveref}

\graphicspath{{figures/}{logos/}}
\newcommand{\illustration}[3]{\IfFileExists{figures/#1}{\includegraphics[width=#2]{#1}}{#3}}

\usepackage{fancyhdr}
\newcommand{\runningtitle}{Do Clinician-Scored Examples Help LLMs Code ADOS-2?}
\newcommand{\crest}[2]{\raisebox{-0.5\height}{\includegraphics[height=#2]{#1}}}
\newcommand{\schoolrow}{%
  \crest{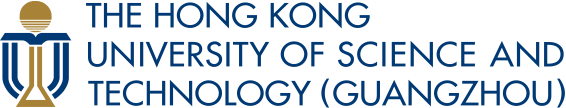}{0.85cm}\hspace{1.6em}%
  \crest{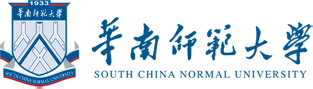}{0.85cm}\hspace{1.6em}%
  \crest{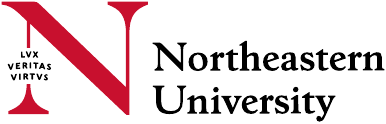}{0.85cm}}
\newcommand{\crestrow}{%
  \crest{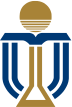}{0.5cm}\hspace{0.5em}\crest{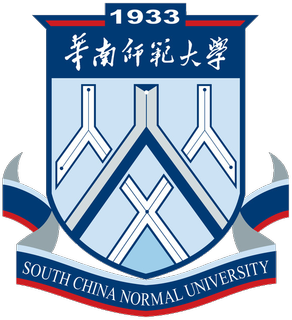}{0.5cm}\hspace{0.5em}%
  \crest{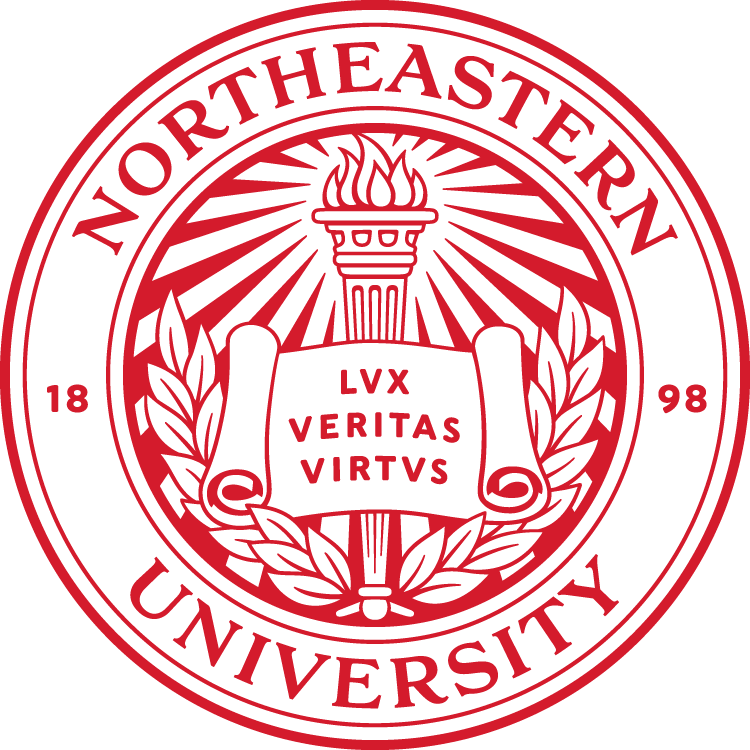}{0.5cm}}
\fancypagestyle{firstpage}{\fancyhf{}%
  \fancyhead[L]{\schoolrow}%
  \fancyfoot[C]{\thepage}%
  }
\fancypagestyle{main}{\fancyhf{}%
  \fancyhead[L]{\crestrow}%
  \fancyhead[R]{\small\itshape\color{black!65}\runningtitle}%
  \fancyfoot[C]{\thepage}%
  }
\begin{document}
\thispagestyle{firstpage}
\begin{center}
\vspace*{-0.4em}
{\color{black}\rule{\textwidth}{1.6pt}}\\[0.7em]
{\LARGE\bfseries Do Clinician-Scored Examples Help LLMs Code ADOS-2\\[0.25em] from Hour-Long Transcripts?\par}
\vspace{0.55em}
{\color{black}\rule{\textwidth}{1.6pt}}\\[1.0em]
{\large Zeteng Lin\textsuperscript{1}\quad Yikang Liu\textsuperscript{2}\quad Bolin Zeng\textsuperscript{2}\quad Mengyuan Ding\textsuperscript{2}\quad Xihui Wu\textsuperscript{1}\\[0.2em]
Kangyou Li\textsuperscript{3}\quad Chengzhe Fu\textsuperscript{4}\quad Xin Wang\textsuperscript{3}}\\[0.5em]
{\small \textsuperscript{1}The Hong Kong University of Science and Technology (Guangzhou), Guangzhou, China}\\
{\small \textsuperscript{2}Northeastern University, USA}\\
\parbox{0.92\textwidth}{\centering\small \textsuperscript{3}Key Laboratory of Brain, Cognition and Education Sciences, Ministry of Education, Institute for Brain Research and Rehabilitation, and Guangdong Key Laboratory of Mental Health and Cognitive Science, South China Normal University, Guangzhou, China}\\
{\small \textsuperscript{4}School of Politics and Public Administration, South China Normal University, Guangzhou, China}%

\end{center}
\vspace{0.4em}

\begin{center}
\begin{minipage}{0.86\textwidth}
\centerline{\large\bfseries Abstract}
\vspace{0.4em}
\small Large language models can apply a written rubric to a transcript, making automated coding of the Autism Diagnostic Observation Schedule (ADOS-2) look within reach. A common way to align such a rater with clinicians is to show it expert-scored examples, but whether correct labels improve agreement is rarely tested. We test this for eight speech-based items of hour-long ADOS-2 Module~3 sessions from 48 children, with one local model and a pre-registered, placebo-controlled design. We compare few-shot scoring, an anchored comparison in which the model only judges a child better, similar or worse than scored anchors and a fixed rule gives the score, post-hoc calibration, protocol-based evidence routing and direct scoring of the full transcript. On 40 held-out children, agreement with clinicians was modest, with quadratic weighted kappa between 0.22 and 0.30. Replacing every example's label with a wrong one changed about a third of the few-shot predictions, more often towards the corrupted labels than chance, but did not lower agreement, 0.25 with correct and 0.23 with wrong labels. A larger second model did the same. The anchored comparison did not improve on few-shot scoring, its cross-child judgements were no more accurate than those implied by absolute scores, and routing did not help. Among the registered, uncalibrated arms, direct scoring of the full transcript agreed best at 0.30. Exploratory follow-up analyses found that the model's graded judgements ranked children better than its final scores, that existing batch calibration raised agreement to 0.46 on the same children, and that in controlled synthetic sessions scoring episodes separately did not beat calibrated whole-transcript scoring. Expert-scored examples do not by themselves calibrate model-based clinical raters, and claims of calibration or added pipeline complexity need corrupted-label controls and strong baselines.

\end{minipage}
\end{center}

\begin{figure}[h]
\centering
\illustration{teaser.png}{\textwidth}{%
  \fbox{\parbox[c][4.4cm][c]{0.96\textwidth}{\centering\small\color{gray}
  Fig.~1 illustration (\texttt{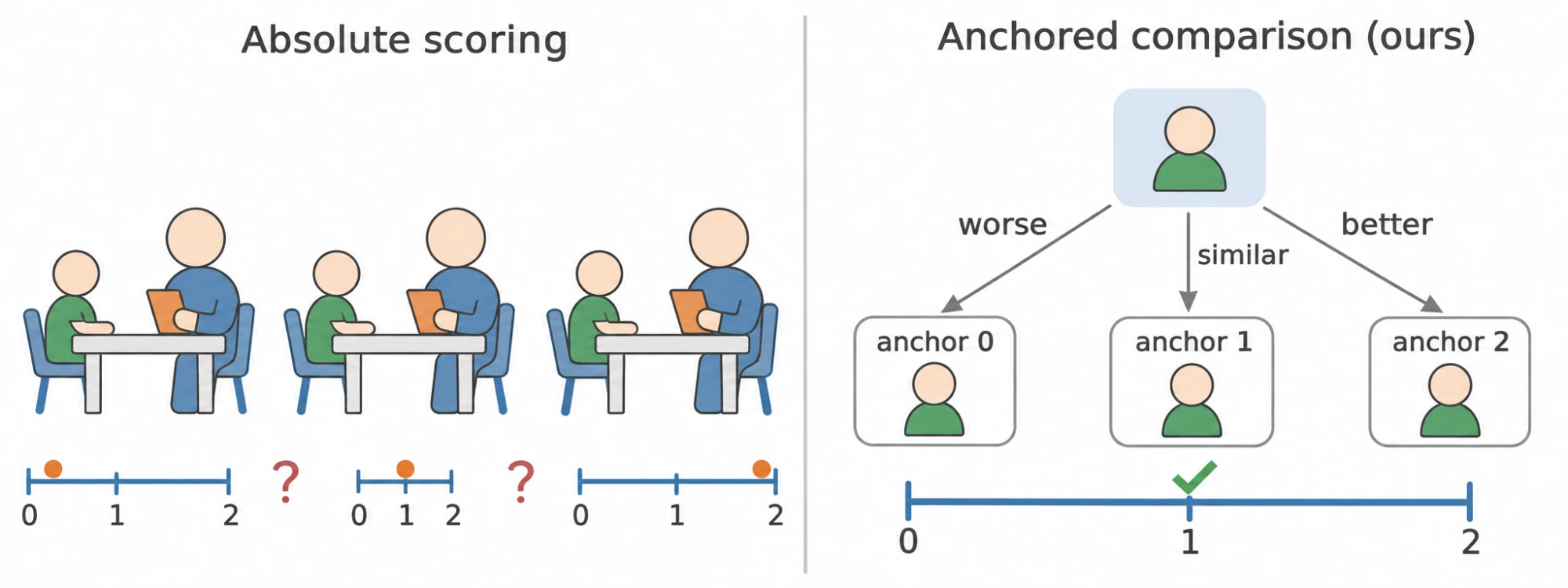}) --- prompt in \texttt{FIGURE\_PROMPTS.md}}}}
\caption{\textbf{Two ways to use clinician-scored children, as tested in this study.} Left: the model scores each child on its own, which in our video pilots produced scores that were not comparable across children. Right: the anchored comparison we designed, in which the model only judges a child as better, similar or worse than clinician-scored anchor children and a fixed rule reads the score off the anchors' labels. The study compares these designs with direct scoring of the full transcript and with placebos in which the anchors carry wrong labels. In the pre-registered evaluation (\cref{sec:results}) the comparison did not improve agreement, and correct anchor labels showed no improvement over corrupted ones.}
\label{fig:teaser}
\end{figure}

\section{Introduction}

The Autism Diagnostic Observation Schedule, Second Edition (ADOS-2) is a semi-structured, standardized observation in which a trained examiner leads a child through a fixed sequence of activities and then codes the child's behaviour on a set of items~\cite{lord2000ados,lord2012ados2}. Each item is a judgement about the whole session, so coding requires a clinician to keep an hour of interaction in mind. Large language models (LLMs) can read a transcript, follow a written rubric and explain a score, which makes computer-assisted item coding look within reach~\cite{jiang2024adoscopilot,hu2025exploiting,prescott2026boscc}. We study this for eight speech-based items of ADOS-2 Module~3, scored from automatic transcripts of hour-long sessions. The task is item coding to support clinicians. It is not autism diagnosis.

A common way to align such a rater with clinicians is to show it examples scored by experts, as few-shot demonstrations~\cite{jiang2024adoscopilot} or as anchors to compare against~\cite{poudel2026anchors}. The assumption is that the examples carry the experts' scale to the model. This assumption is rarely tested directly. A model may change its predictions when it sees labelled examples, yet that does not show that correct labels make its scores agree better with clinicians. Without a control in which the labels are wrong, label sensitivity and calibration benefit cannot be told apart. The question matters in our setting because, in pilot experiments on the same 48 sessions (\cref{sec:challenge}), model judgements ordered behaviour within a child but did not place children on a common scale, and fewer than fifty labelled sessions are too few to learn that scale by fine-tuning.

We therefore test whether clinician-scored examples transmit the clinician's scale, and whether the form in which they are given matters (\cref{fig:teaser}). We compare few-shot scoring with the examples' scores shown, an anchored comparison in which the model only states whether the target child shows fewer, similar or more difficulties than each anchor and a fixed rule turns these relations into a score, post-hoc calibration of absolute scores, and direct scoring of the full transcript. Each ingredient has a placebo: anchors with deliberately wrong labels, evidence from random instead of protocol-routed windows, and relations implied by the model's own absolute scores. All arms use the same local model. The hypotheses, split, metrics and pass criteria were registered before any scoring call, and a later amendment repeated the key tests with a second, larger model.

The main results are largely negative. On 40 held-out children, agreement with clinicians was modest for every arm (mean quadratic weighted kappa 0.22--0.30). Replacing every example's label with a wrong one changed about a third of the few-shot predictions, more often towards the corrupted labels than chance would produce, but did not lower agreement (0.25 with correct labels, 0.23 with wrong ones). The second model behaved the same way. The anchored comparison did not improve on few-shot scoring, and among the registered arms, which were not calibrated, direct scoring of the full transcript agreed best with clinicians (0.30). Follow-up analyses, which were not part of the registration, then asked where scoring hour-long sessions goes wrong. They found that the model's graded judgements ranked children better than its final scores, that existing batch calibration raised agreement to 0.46 on the same children, and that splitting sessions into separately scored episodes did not beat calibrated whole-transcript scoring in controlled synthetic sessions. On this data, then, the examples moved the model's scores without bringing them closer to the clinicians', and the extra processing steps we tried lost some of what a reading of the whole session retains.

Our contributions are:
\begin{itemize}
\item \textbf{A controlled evaluation of expert-scored examples.} A pre-registered, placebo-controlled test of whether clinician-scored examples improve an LLM's agreement with clinicians on hour-long ADOS-2 transcripts, replicated with a second model.
\item \textbf{A separation of label sensitivity from calibration benefit.} Corrupting the expert labels changed the model's predictions without lowering its agreement with clinicians, and anchored comparison did not improve on simpler scoring designs.
\item \textbf{An exploratory analysis of long-session scoring.} Analyses of the probability read-out, controlled stress tests with mixed evidence, and episode-level decomposition identify conditions in which added pipeline structure loses information compared with calibrated whole-transcript scoring.
\end{itemize}

\section{Related Work}

\paragraph{Automated autism assessment.}
Computer vision for autism research has been reviewed systematically~\cite{debelen2020computer}, as has AI-based diagnosis from face, voice and text~\cite{mohammadi2025scoping}. Closest to our setting, ADOS-Copilot scores and explains ADOS-2 items with LLMs on real clinical recordings~\cite{jiang2024adoscopilot}. Zero-shot LLMs have been applied to ADOS-2 Module~4 audio transcripts to detect autism-associated language disorders~\cite{hu2025exploiting}, and a multi-agent dialogue system selects questioning strategies to elicit social-language traits in Module~4~\cite{hu2026tpa}. General-purpose LLMs predict speech-related codes of the BOSCC observation instrument from transcripts, diarized transcripts or audio~\cite{prescott2026boscc}, and speech, video and text foundation models have been benchmarked on child-inclusive autism diagnostic videos, succeeding on coarse tasks such as activity recognition and struggling on fine-grained ones~\cite{kommineni2025interspeech}. CARE-VL fine-tunes a vision-language model on expert-marked query--response intervals and aggregates clip-level outputs with an LLM into an autism/typical-development decision~\cite{yoo2025carevl}. These systems produce absolute scores or classifications, in some cases with few-shot examples. None of them places a child relative to clinician-scored anchor children through comparisons.

\paragraph{LLMs as rubric raters of long sessions.}
Outside autism, ChatGPT has been used zero-shot to score classroom transcript segments against observation instruments~\cite{wang2023teachercoach}, and language models have been applied to high-inference measures of instruction quality from long, noisy classroom transcripts~\cite{xu2024promises}. In depression assessment, LLMs predict PHQ-8 items from interview transcripts~\cite{teferra2025depression} and produce PHQ-8-aligned symptom summaries for explainable recognition~\cite{zheng2025explainmdrc}. LLM judging more broadly has documented position, verbosity and self-enhancement biases~\cite{zheng2023judging,wang2024fair}, and G-Eval scores text with chain-of-thought and form filling~\cite{liu2023geval}.

\paragraph{Comparative and anchored judging.}
Comparative judgement has a long history in psychometrics~\cite{thurstone1927law,bradley1952rank}. For LLMs, pairwise comparison has been shown to be an effective zero-shot evaluator~\cite{liusie2024comparative}, with efficient aggregation of comparisons~\cite{liusie2024efficient}, pairwise ranking prompting for retrieval~\cite{qin2024prp}, and pairwise comparisons converted into essay scores~\cite{shibata2025lces}. Most relevant, Poudel et~al.\ compare few-shot direct scoring with anchor-based comparative grading of essays using human-scored anchors and find the comparative approach more stable~\cite{poudel2026anchors}. We adopt the anchor mechanism and study what it takes to apply it to hour-long clinical sessions with many activities, which involves protocol routing, score-free evidence and a placebo-controlled test of each module.

\section{Data and Setting}
\label{sec:data}

\paragraph{Sessions.}
The dataset contains 50 ADOS-2 Module~3 sessions, 38.3 hours in total, each 28 to 84 minutes long, recorded at $1920\times1080$ and 25 frames per second by a fixed wide-angle camera mounted high in a corner of the room. Two sessions are byte-identical recordings registered with different total scores (2 and 11). Because the correct label cannot be determined, both are excluded, leaving $n=48$ children. The sessions arrived in two batches whose mean ADOS totals differ (6.06 for the first 16 children versus 10.50 for the remaining 32, Mann--Whitney $p=0.017$), so we also report results within each batch.

\paragraph{Items.}
Clinicians provided algorithm-level item scores (0, 1 or 2, with a raw score of 3 mapped to 2 under the scoring rules). We study the eight items whose evidence is primarily verbal (\cref{tab:items}): A4 stereotyped or idiosyncratic use of words or phrases, A7 reporting of events, A8 conversation, B7 quality of social overtures, B9 quality of social response, B10 amount of reciprocal social communication, B11 overall quality of rapport, and D4 excessive interest in unusual or highly specific topics or objects, or repetitive behaviours. Items that require seeing the child (eye contact, directed facial expressions, gestures, shared enjoyment, sensory interests, hand mannerisms) are out of scope, because the child mostly faces away from the camera, a face detector found a face in 1\% of sampled frames, and head keypoint confidences were 0.005--0.28.

\begin{table}[t]
\centering
\caption{Items studied ($n=48$ children): counts per algorithm score, accuracy of always predicting the most frequent score, and the windows each item reads.}
\label{tab:items}
\small
\setlength{\tabcolsep}{4pt}
\begin{tabular}{llrrrrl}
\toprule
Item & Short name & 0 & 1 & 2 & Maj. & Input \\
\midrule
A4 & Stereotyped language & 36 & 8 & 4 & 75\% & all \\
A7 & Reporting events & 24 & 14 & 10 & 50\% & routed \\
A8 & Conversation & 19 & 19 & 10 & 40\% & routed \\
B7 & Social overtures & 10 & 28 & 10 & 58\% & routed \\
B9 & Social response & 21 & 22 & 5 & 46\% & all \\
B10 & Reciprocal communication & 20 & 16 & 12 & 42\% & all \\
B11 & Rapport & 18 & 22 & 8 & 46\% & all \\
D4 & Restricted interests & 32 & 10 & 6 & 67\% & all \\
\bottomrule
\end{tabular}
\end{table}

\begin{figure}[t]
\centering
\includegraphics[width=\textwidth]{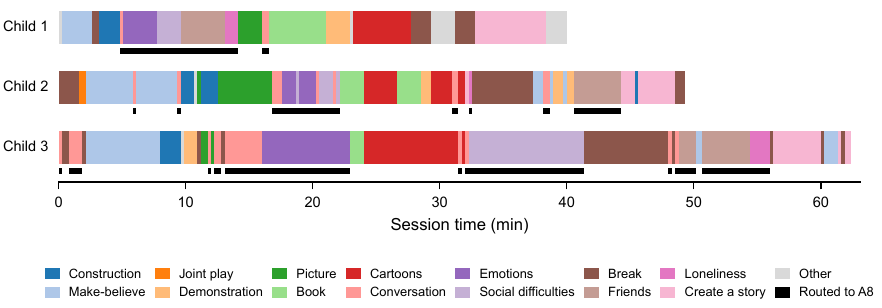}
\caption{\textbf{Protocol timelines of three children} (automatic activity labels per 16-second window). Black ticks mark the windows routed to item A8 (conversation), namely the conversation, emotions, social-difficulties, friendship and loneliness activities. Sessions differ in length, order and duration of activities, so a fixed time window cannot stand in for the protocol.}
\label{fig:timeline}
\end{figure}

\paragraph{Transcripts and protocol timeline.}
Each session was transcribed by an automatic speech recogniser (Qwen3-ASR~\cite{shi2026qwen3asr}) in consecutive 16-second windows without speaker labels. Transcripts average about 8,400 Chinese characters per session. Each window was tagged with one of 14 Module~3 activities or ``other'' by an LLM reading the transcript (\cref{fig:timeline}). Against a shuffle control, contiguous activity blocks were a median of 5.4 times longer than under shuffled labels, and all 50 sessions exceeded a pre-set threshold of three times.

\section{Pilot Evidence of Cross-Subject Inconsistency}
\label{sec:challenge}

\begin{figure}[t]
\centering
\includegraphics[width=\textwidth]{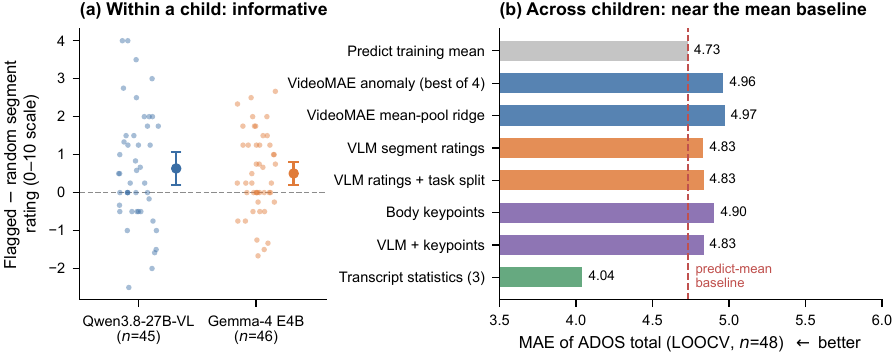}
\caption{\textbf{Pilot evidence.} (a) Within a child, a blind judge rates segments flagged by the system above random segments of the same child (each dot is one child, error bars are bootstrap 95\% intervals of the mean). The effect holds for two unrelated judge models. (b) Across children, every video- or vision-derived predictor of the ADOS total is at or above the error of always predicting the training mean (dashed). Only simple transcript statistics improve on it.}
\label{fig:challenge}
\end{figure}

Before the main study we ran four rounds of pilot experiments, each with pass criteria fixed in advance. They used video segments and ADOS total scores, not item-level transcript scoring, so they motivate the question of the main study but do not test it. \Cref{fig:challenge} summarises them, and \cref{app:pilot} gives the details.

Within a child, model judgements carried signal. A blind vision-language judge rated the segments our system flagged as most atypical 0.64 points higher on a 0--10 scale than random segments of the same child (Wilcoxon $p=0.0106$, $n=45$), and an unrelated, much smaller judge model preserved the effect ($p=0.0044$). Across children, the same kind of judgement did not transfer. No video-based predictor of the ADOS total beat always predicting the training mean (MAE 4.73 versus 4.96--5.63), and per-segment ratings by a vision-language model, aggregated per child, did no better (4.83--4.85). Learned segment features identified the source recording with 79--85\% accuracy, which suggests that cross-child comparisons mostly reflected who and where, and less the behaviour.

These pilots led us to ask whether clinician-scored reference children can supply the cross-child scale that the segment-level judgements lacked, and to design the anchored comparison of \cref{sec:method} as one way of doing so. Because the pilots concerned segment-level judgements and total scores, they should be read as motivation and not as evidence about item-level transcript scoring.

\section{Anchored Comparative Rating}
\label{sec:method}

\begin{figure}[t]
\centering
\resizebox{\textwidth}{!}{
\definecolor{mBlue}{HTML}{3A6EA5}\definecolor{mOrange}{HTML}{E07B39}\definecolor{mGreen}{HTML}{4C9A6A}
\definecolor{mPurple}{HTML}{8E7CC3}\definecolor{mLight}{HTML}{BFD3EA}\definecolor{mGrey}{HTML}{DADADA}
\definecolor{mCard}{HTML}{E8F0F8}
\begin{tikzpicture}[x=1cm,y=1cm,
  head/.style={font=\sffamily\bfseries\small, anchor=base},
  arr/.style={-{Stealth[length=2.2mm]}, line width=0.7pt, black!65},
  card/.style={draw=black!55, rounded corners=2pt, line width=0.5pt, fill=white},
  stamp/.style={fill=mCard, rounded corners=1.5pt, font=\sffamily\scriptsize, inner sep=1.6pt, minimum width=8mm},
  line/.style={line width=1.4pt, black!35, line cap=round},
  rel/.style={font=\sffamily\scriptsize, inner sep=1pt}]
\node[head] at (1.45,3.0) {Hour-long session};
\node[head] at (5.1,3.0) {Protocol routing};
\node[head] at (8.75,3.0) {Score-free evidence};
\node[head] at (12.55,3.0) {Compare to anchors};
\node[head] at (16.35,3.0) {Fixed rule};
\foreach \x/\w/\c in {0/0.45/mLight,0.45/0.85/mOrange,1.3/0.5/mGreen,1.8/0.5/mPurple,2.3/0.6/mLight}
  \fill[\c] (\x,1.2) rectangle ++(\w,0.4);
\draw[black!55, rounded corners=1pt, line width=0.5pt] (0,1.2) rectangle (2.9,1.6);
\draw[black!60, line width=0.6pt] (1.0,2.2) circle (0.28); \draw[black!60, line width=0.6pt] (1.0,2.2) -- (1.0,2.38) (1.0,2.2) -- (1.13,2.2);
\draw[black!60, line width=0.6pt, rounded corners=1.5pt] (1.82,2.07) rectangle (2.02,2.47);
\draw[black!60, line width=0.6pt] (1.72,2.2) arc (180:360:0.2) (1.92,2.0) -- (1.92,1.9) (1.8,1.9) -- (2.04,1.9);
\draw[arr] (3.2,1.4) -- (3.75,1.4);
\foreach \x/\c in {3.95/mGrey,4.4/mBlue,4.85/mGrey,5.3/mBlue,5.75/mGrey}
  \fill[\c] (\x,1.2) rectangle ++(0.45,0.4);
\draw[black!55, rounded corners=1pt, line width=0.5pt] (3.95,1.2) rectangle (6.2,1.6);
\draw[black!60, line width=0.6pt, line join=round] (4.78,2.5) -- (5.42,2.5) -- (5.17,2.15) -- (5.17,1.92) -- (5.03,1.92) -- (5.03,2.15) -- cycle;
\draw[arr] (6.5,1.4) -- (7.05,1.4);
\draw[card] (7.75,0.15) rectangle (9.75,2.6);
\foreach \y/\t in {2.25/00:05,1.7/12:18,1.15/34:07,0.6/55:30}{
  \node[stamp] at (8.25,\y) {\t};
  \draw[line] (8.75,\y+0.07) -- (9.55,\y+0.07); \draw[line] (8.75,\y-0.09) -- (9.3,\y-0.09);}
\draw[black!60, line width=0.6pt] (8.75,2.88) circle (0.0); 
\draw[arr] (10.0,1.4) -- (10.55,1.4);
\draw[card] (10.75,0.15) rectangle (12.45,2.6);
\foreach \y/\t in {2.25/00:05,1.7/12:18,1.15/34:07,0.6/55:30}{
  \node[stamp] at (11.25,\y) {\t};
  \draw[line] (11.75,\y+0.07) -- (12.3,\y+0.07); \draw[line] (11.75,\y-0.09) -- (12.1,\y-0.09);}
\foreach \y/\c in {2.2/mGreen,1.4/mOrange,0.6/mBlue}{
  \draw[card] (13.85,\y-0.34) rectangle (14.55,\y+0.34);
  \fill[\c] (14.2,\y+0.1) circle (0.12);
  \fill[\c] (14.0,\y-0.26) arc (180:0:0.2) -- cycle;}
\draw[arr] (12.6,1.55) -- (13.75,2.2) node[rel, pos=0.5, above, sloped] {better};
\draw[arr] (12.6,1.4) -- (13.75,1.4) node[rel, pos=0.5, above] {similar};
\draw[arr] (12.6,1.25) -- (13.75,0.6) node[rel, pos=0.5, below, sloped] {worse};
\draw[arr] (14.75,1.4) -- (15.3,1.4);
\draw[black!60, line width=0.7pt] (15.5,2.4) -- (17.2,2.4); \draw[black!60, fill=white, line width=0.6pt] (16.35,2.4) circle (0.09);
\foreach \x/\l in {15.6/0,16.35/1,17.1/2}{
  \draw[black!50, line width=0.5pt] (\x,2.4) -- (\x-0.2,1.85) (\x,2.4) -- (\x+0.2,1.85);
  \node[draw=mBlue!70, fill=mCard, rounded corners=2pt, font=\sffamily\bfseries\small, minimum width=6.5mm, minimum height=4.6mm, inner sep=1pt] at (\x,1.7) {\l};}
\draw[arr] (16.35,1.4) -- (16.35,1.0);
\node[draw=mBlue!70, fill=mCard, rounded corners=4pt, minimum width=12mm, minimum height=6.5mm] at (16.35,0.6) {};
\draw[mGreen, line width=1.6pt, line cap=round, line join=round] (16.08,0.62) -- (16.26,0.44) -- (16.62,0.8);
\end{tikzpicture}}
\caption{\textbf{Anchored comparative rating.} For one item, the LLM extracts score-free evidence from the protocol activities where the item is observable, then judges the target child as better, similar or worse than each anchor child without seeing the anchors' scores. The score is computed from these relations and the anchors' clinician labels by a fixed rule (\cref{eq:agg}).}
\label{fig:method}
\end{figure}
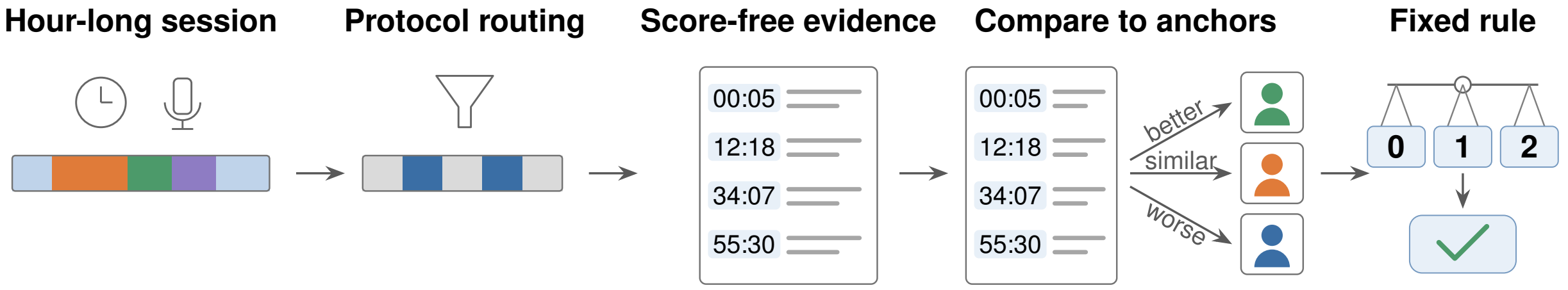

Let $s$ be a target session, $i$ an item with levels $\mathcal{L}=\{0,1,2\}$, and $\mathcal{P}$ the pool of other labelled sessions. The method has four modules (\cref{fig:method}).

\paragraph{M1: Protocol alignment.} Each 16-second transcript window carries an activity label from the session timeline (\cref{sec:data}).

\paragraph{M2: Item routing.} Each item is mapped to the activities where its evidence is expected. For A7 the scoring instructions name the demonstration task explicitly, and for A8 and B7 we derived the activities from the observation notes of the Module~3 protocol before any scoring call. A7 is routed to the demonstration task, storybook, cartoons and conversation activities, A8 to conversation and the emotions, social-difficulties, friendship and loneliness interviews, and B7 to construction, make-believe play and the break. The remaining five items are defined over the whole session and receive all windows.

\paragraph{M3: Score-free evidence.} For each session and item, the LLM reads the routed windows and lists up to eight pieces of evidence about the child, each with a timestamp, a short verbatim quote and an objective note, followed by a two-to-three-sentence summary. The prompt forbids scores and evaluative words, and asks the model to infer from content which utterances are the child's, since transcripts carry no speaker labels.

\paragraph{M4: Anchored comparison.} For each level $\ell\in\mathcal{L}$ one anchor child $a_\ell\in\mathcal{P}$ with score $\ell$ on item $i$ is drawn deterministically (by a hash of target, item and level). The LLM sees the item's full rubric, the M3 evidence of the three anchors \emph{without their scores} in a fixed random order, and the target's evidence, and outputs only a relation $r_\ell\in\{-1,0,+1\}$ per anchor: the target shows fewer difficulties than, about the same as, or more difficulties than the anchor on this item. The score is
\begin{equation}
\hat{y}(s,i)=\arg\max_{y\in\mathcal{L}}\ \prod_{\ell\in\mathcal{L}} \begin{cases}0.7 & r_\ell=\operatorname{sign}(y-\ell)\\ 0.15 & \text{otherwise,}\end{cases}
\label{eq:agg}
\end{equation}
with ties broken towards the median label of $\mathcal{P}$. The rule is fixed in advance and has no fitted parameters. The clinician's scale enters only through the anchors' labels.

The model never states a score in M4. The design rationale was to separate two roles that absolute scoring mixes. The model judges which of two children shows more of a behaviour, and the anchors' labels supply the scale. Whether the first role is in fact performed more reliably than absolute scoring is tested directly by H6.

\section{Experimental Design}
\label{sec:design}

\paragraph{Model.} All arms use the same locally served LLM, Qwen3.8-27B from the Qwen3 family~\cite{yang2025qwen3}, under vLLM~\cite{kwon2023vllm}, at temperature 0, with reasoning enabled for scoring and comparison calls and disabled for the descriptive extraction step. No recording or transcript leaves the research team's servers.

\paragraph{Arms.} We compare the method (A) with baselines and placebos that each remove or corrupt one ingredient:
\begin{itemize}
\item \textbf{D} (direct): full transcript and item rubric, and the model outputs a raw score.
\item \textbf{Z} (absolute on evidence): M1--M3, then the model scores the evidence against the rubric.
\item \textbf{Zcal}: Z mapped through leave-one-out calibration with the same labels (for each raw score, the most frequent true score among the other children with that raw score).
\item \textbf{F} (few-shot): the same three anchors shown \emph{with} their scores, and the model outputs the score. This is the direct condition of \citet{poudel2026anchors} and a few-shot setting as in \citet{jiang2024adoscopilot}.
\item \textbf{PF}: F with every anchor label cyclically shifted ($0\to1\to2\to0$).
\item \textbf{PA}: A's relations aggregated with the shifted labels.
\item \textbf{R}: A with evidence for target and anchors extracted from the same number of randomly chosen windows, for the three routed items.
\end{itemize}

\paragraph{Split and metrics.} Eight children, chosen by stratifying on the total score before any scoring call, were used only to check output formats. Results are reported on the remaining 40, with all 48 as a secondary analysis. Anchors for a target are always other children. Agreement is measured by quadratic weighted kappa (QWK)~\cite{cohen1968weighted} per item, and the primary endpoint is mean QWK over the eight items. Paired differences between arms use a subject-level bootstrap (10,000 resamples) with one-sided 95\% bounds~\cite{efron1979bootstrap}. The absolute test of A uses a permutation null that shuffles children's label vectors jointly across items. Unparseable outputs are replaced by the leave-one-out majority score and counted.

\paragraph{Hypotheses.} H1: A agrees with clinicians above chance and beats direct scoring D. H2: A beats few-shot scoring F with the same anchors. H3: the anchor labels are causal (A $>$ PA and F $>$ PF). H4: protocol routing is causal (A $>$ R on routed items). H5: A beats post-hoc calibration Zcal with the same labels. H2--H5 are Holm-corrected as a family~\cite{holm1979}. Prompts were not tuned on any child's agreement.

\paragraph{Mechanism test.} H1--H5 test whether the method works, and H6 tests why. For every (target, item, anchor) triple the true relation is known from the clinician labels, $r^*=\operatorname{sign}(y_{\text{target}}-\ell)$. We compare the accuracy of the relations the model states in A with the relations implied by its own absolute scores in Z, $\operatorname{sign}(\hat z_{\text{target}}-\hat z_{\text{anchor}})$, i.e.\ scoring each child on the model's scale and then comparing. H6a: A's relation accuracy exceeds chance (permutation of children's labels). H6b: it exceeds the Z-implied accuracy (subject-level bootstrap). If H6b fails, any end-to-end gain cannot be attributed to more reliable relative judgement. H6 was added to the pre-registration before any evaluation was run and is reported separately from the H2--H5 family.

\paragraph{Decoding and model scale.} To ask whether ignoring the anchor labels is a limitation of a 27B model, a further amendment, registered after the main results and before any call to the second model, re-runs arms Z, F, PF and A with DeepSeek-V4.1-Flash, run locally on the team's GPU cluster. On that hardware the model can only be run without reasoning, so the amendment uses \emph{forced-choice} decoding. The same prompts are followed by a fixed answer prefix and the answer is the most probable allowed option (score digit, or better/similar/worse for each anchor). Qwen3.8-27B is re-run in the identical mode as the matched control, so that model and decoding are not confounded. The registered tests are, for the larger model, F $>$ PF (labels used, H7) and A $>$ F (H8), Holm-corrected. The same differences are reported for the Qwen control without pass/fail.

\section{Results}
\label{sec:results}
\IfFileExists{results_tables.tex}{\begin{table}[t]
\centering
\caption{Agreement with clinicians on the test set ($n=40$ children): quadratic weighted kappa (QWK) per item and mean over items, mean exact agreement, and share of outputs that could not be scored (replaced by the leave-one-out majority). $^\dagger$R covers the three routed items only; its mean is over those items. Majority (LOO) predicts for each child the most frequent score among the other children. Leaving the child out tips the majority away from the child's own score whenever the classes are nearly balanced, so this baseline can agree worse than chance (negative QWK).}
\label{tab:main}
\small
\setlength{\tabcolsep}{3.6pt}
\begin{tabular}{lrrrrrrrrrrr}
\toprule
Arm & Mean & A4 & A7 & A8 & B7 & B9 & B10 & B11 & D4 & Exact & NA \\
\midrule
Majority (LOO) & \textbf{$-$0.14} & 0.00 & 0.00 & $-$0.63 & 0.00 & $-$0.46 & 0.00 & 0.00 & 0.00 & 42\% & 0.0\% \\
D: direct, full transcript & \textbf{0.30} & 0.21 & 0.25 & 0.21 & 0.11 & 0.08 & 0.48 & 0.51 & 0.57 & 45\% & 1.2\% \\
Z: absolute on evidence & \textbf{0.22} & 0.06 & 0.13 & 0.48 & 0.08 & 0.17 & 0.36 & 0.19 & 0.24 & 37\% & 0.6\% \\
Zcal: Z + LOO calibration & \textbf{0.06} & 0.00 & 0.00 & $-$0.06 & 0.00 & 0.17 & 0.18 & 0.17 & 0.00 & 49\% & 0.9\% \\
F: few-shot with anchors & \textbf{0.25} & 0.10 & 0.28 & 0.46 & 0.17 & 0.10 & 0.47 & 0.19 & 0.21 & 38\% & 0.6\% \\
PF: F, wrong labels & \textbf{0.23} & 0.07 & 0.25 & 0.36 & 0.23 & 0.26 & 0.30 & 0.26 & 0.14 & 40\% & 0.6\% \\
A: anchored comparison & \textbf{0.22} & 0.35 & 0.06 & 0.33 & 0.12 & 0.11 & 0.32 & 0.22 & 0.22 & 45\% & 0.3\% \\
PA: A, wrong labels & \textbf{0.15} & 0.12 & 0.24 & $-$0.14 & 0.11 & 0.31 & 0.19 & 0.17 & 0.18 & 43\% & 0.3\% \\
R: A, random windows$^\dagger$ & \textbf{0.22} & -- & $-$0.05 & 0.42 & 0.28 & -- & -- & -- & -- & 43\% & 0.0\% \\
\bottomrule
\end{tabular}
\end{table}

\begin{table}[t]
\centering
\caption{Pre-registered paired tests on the test set: difference in mean QWK (H6b: in relation accuracy), one-sided 95\% bootstrap lower bound, one-sided $p$, and verdict. H2--H5 are Holm-corrected; H3 requires H3a and H3b. Differences are computed from unrounded estimates, and displayed values are rounded. Permutation test of A against chance: mean QWK 0.22, $p=0.0002$. H6a: A's relations above chance, permutation $p=0.0049$.}
\label{tab:tests}
\small
\begin{tabular}{llrrrc}
\toprule
& Comparison & $\Delta$ & Lower 95\% & $p$ & Pass \\
\midrule
H1 & A vs D & $-$0.08 & $-$0.19 & 0.898 & no \\
H2 & A vs F & $-$0.03 & $-$0.10 & 0.774 & no \\
H3a & A vs PA & 0.07 & $-$0.06 & 0.222 & no \\
H3b & F vs PF & 0.01 & $-$0.08 & 0.419 & no \\
H4 & A vs R (routed items) & $-$0.05 & $-$0.20 & 0.684 & no \\
H5 & A vs Zcal & 0.16 & 0.06 & 0.004 & yes \\
\midrule
H6b & relation acc.\ A vs Z-implied & 0.01 & $-$0.02 & 0.242 & no \\
\bottomrule
\end{tabular}
\end{table}

\begin{table}[t]
\centering
\caption{Mechanism test (H6): accuracy of cross-child relations (better / similar / worse) over 960 (target, item, anchor) triples on the test set, overall and on triples whose true relation is not ``similar''.}
\label{tab:relations}
\small
\begin{tabular}{lrr}
\toprule
Source of the relation & Accuracy & Ordering only \\
\midrule
A: stated relation & 46.0\% & 52.0\% \\
Z-implied: $\mathrm{sign}(\hat z_t-\hat z_a)$ & 44.7\% & 40.9\% \\
Z vs anchor label: $\mathrm{sign}(\hat z_t-\ell)$ & 52.9\% & 60.9\% \\
F vs anchor label: $\mathrm{sign}(\hat f_t-\ell)$ & 55.8\% & 64.7\% \\
\bottomrule
\end{tabular}
\end{table}

\begin{table}[t]
\centering
\caption{Secondary analyses: mean QWK on all 48 children; within each recording batch on the test set (batch 1: earlier sessions, batch 2: later sessions); Spearman correlation between predicted and clinician sums over the eight items on the test set.}
\label{tab:secondary}
\small
\begin{tabular}{lrrrr}
\toprule
Arm & All 48 & Batch 1 & Batch 2 & Sum $\rho$ \\
\midrule
Majority (LOO) & $-$0.14 & $-$0.18 & $-$0.09 & $-$0.69 \\
D: direct, full transcript & 0.34 & 0.22 & 0.27 & 0.60 \\
Z: absolute on evidence & 0.24 & 0.11 & 0.22 & 0.61 \\
Zcal: Z + LOO calibration & 0.08 & $-$0.01 & 0.06 & 0.31 \\
F: few-shot with anchors & 0.27 & 0.07 & 0.26 & 0.57 \\
PF: F, wrong labels & 0.27 & 0.09 & 0.22 & 0.53 \\
A: anchored comparison & 0.24 & 0.09 & 0.23 & 0.53 \\
PA: A, wrong labels & 0.17 & 0.07 & 0.17 & 0.25 \\
\bottomrule
\end{tabular}
\end{table}
}{}

\begin{figure}[t]
\centering
\includegraphics[width=\textwidth]{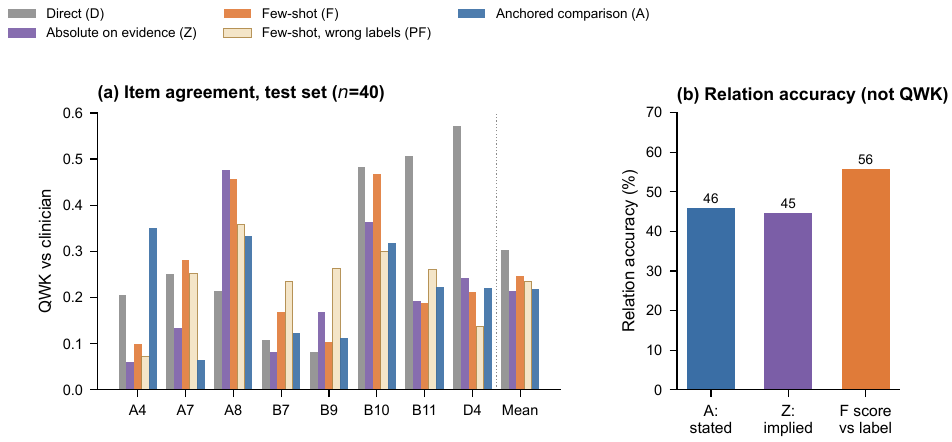}
\caption{\textbf{Pre-registered results on the test set.} (a) Quadratic weighted kappa per item. Direct scoring of the full transcript (D) has the highest mean agreement, mainly because of the whole-session items B10, B11 and D4. Few-shot scoring with deliberately wrong anchor labels (PF) does no worse than with the correct labels (F). (b) A different metric, the percentage of (test child, anchor) pairs whose better/similar/worse relation is correct. ``A: stated'' is the relation the model states directly in the anchored comparison. ``Z: implied'' compares the model's absolute scores of the two children. ``F score vs label'' compares the model's few-shot score of the test child with the anchor's clinician label. Stated relations are no more accurate than implied ones and less accurate than the score-versus-label comparison.}
\label{fig:results}
\end{figure}

\paragraph{Main comparison.}
On the 40 test children, the anchored comparison (A) agreed with clinicians above chance (mean QWK 0.22, permutation $p=0.0002$, \cref{tab:main}). Direct scoring of the full transcript (D) reached 0.30, the highest of the registered arms, none of which was calibrated. H1 required A to exceed D and did not pass ($\Delta=-0.08$, one-sided $p=0.90$, \cref{tab:tests}). Absolute scoring of the extracted evidence (Z) reached 0.22 and few-shot scoring with the anchors (F) 0.25. H2, the registered test of A against F, did not pass either ($\Delta=-0.03$, $p=0.77$).

\paragraph{Correct versus corrupted labels.}
With every anchor label corrupted, few-shot agreement was almost the same as with correct labels (F 0.25 versus PF 0.23). Aggregating A's relations with corrupted labels lowered agreement by 0.07, which was not significant (PA 0.15). H3 required both effects and did not pass. With correct clinician labels, the model's scores did not show better agreement with the clinicians. A non-significant difference is not proof of no effect, but the observed differences were small (F $-$ PF $=+0.01$, one-sided lower 95\% bound $-0.08$).

\paragraph{Sensitivity to the labels (post hoc).}
A lack of benefit does not mean the labels were ignored. We therefore compared F and PF child by child, using the saved predictions and no further model calls. Corrupting the labels changed the predicted level for 107 of 318 (child, item) units, or 34\%. For comparison, the few-shot predictions of two unrelated children for the same item differed in 58\% of units. Of the changed units, 61\% moved exactly to the level that the cyclic label shift points to, against 50\% if the direction were random (binomial $p=0.017$). In the forced-choice runs, which use a single forward pass and so have no reasoning variability, 14\% of predictions changed and 67\% of those moved in the shifted direction ($p=0.013$). So the model did respond to the labels and partly followed them, but the changes cancelled out. Switching from correct to corrupted labels made 45 predictions correct and 38 incorrect. For the reasoning runs there is a caveat. We did not measure how often a repeated run with identical labels changes its prediction, so part of the 34\% may be run-to-run variation. The forced-choice result does not have this problem.

\paragraph{Protocol routing.}
On the three routed items, evidence taken from the same number of randomly chosen windows gave slightly higher mean agreement than protocol routing (R 0.22 versus A 0.17 on those items). H4 did not pass.

\paragraph{Calibration baseline.}
A beat leave-one-out calibration of Z (H5 passed, $\Delta=+0.16$, $p=0.004$). The calibrated arm, however, scored below uncalibrated Z (0.06 versus 0.22). With few children per raw score, the per-value majority mapping often collapsed to a single label, giving QWK 0.00 on four items. H5 therefore says that this calibration rule is a weak baseline. It does not show that anchoring adds information.

\paragraph{Mechanism test.}
Over 960 (target, item, anchor) triples, the relations the model stated in A were correct 46\% of the time. This is above chance (permutation $p=0.005$) but no better than the relations implied by its own absolute scores (45\%, \cref{tab:relations}), so H6b did not pass. Comparing the model's absolute score with the anchor's true label was more accurate (Z 53\%, F 56\%), and more so on triples whose true relation was not ``similar'' (A 52\%, F 65\%).

\paragraph{Where direct reading wins (exploratory).}
These comparisons were not pre-registered. Direct scoring beat absolute scoring of the extracted evidence under the same rubric (D $-$ Z $=+0.09$, bootstrap lower 95\% bound $+0.01$, $p=0.03$). The whole difference came from the five items defined over the whole session (A4, B9, B10, B11, D4), where D reached 0.37 and Z 0.21 ($+0.16$, $p=0.0005$). On the three routed items the two did not differ (D 0.19, Z 0.23). Reducing an hour of interaction to a few evidence snippets seems to lose information that session-wide judgements need, while it costs nothing for items tied to particular activities.

\begin{table}[t]
\centering
\caption{Forced-choice decoding (no reasoning; the answer is the most probable allowed option after a fixed answer prefix), same prompts and anchors as the main arms, test set ($n=40$). Mean and per-item QWK.}
\label{tab:forced}
\small
\setlength{\tabcolsep}{3.6pt}
\begin{tabular}{llrrrrrrrrr}
\toprule
Model & Arm & Mean & A4 & A7 & A8 & B7 & B9 & B10 & B11 & D4 \\
\midrule
Qwen3.8-27B & Z & \textbf{0.26} & 0.09 & 0.29 & 0.42 & 0.20 & 0.22 & 0.39 & 0.31 & 0.18 \\
 & F & \textbf{0.23} & 0.06 & 0.20 & 0.42 & 0.17 & 0.25 & 0.46 & 0.31 & $-$0.01 \\
 & PF (wrong labels) & \textbf{0.24} & 0.06 & 0.17 & 0.40 & 0.11 & 0.40 & 0.27 & 0.25 & 0.25 \\
 & A & \textbf{0.24} & 0.29 & 0.25 & 0.20 & 0.06 & 0.14 & 0.19 & 0.38 & 0.37 \\
 & PA (wrong labels) & \textbf{0.12} & 0.13 & 0.23 & $-$0.06 & 0.30 & $-$0.03 & 0.11 & 0.01 & 0.29 \\
\midrule
DeepSeek-V4.1-Flash & Z & \textbf{0.22} & 0.04 & 0.13 & 0.43 & 0.16 & 0.30 & 0.31 & 0.24 & 0.15 \\
 & F & \textbf{0.26} & 0.09 & 0.23 & 0.34 & 0.17 & 0.47 & 0.35 & 0.27 & 0.15 \\
 & PF (wrong labels) & \textbf{0.26} & 0.07 & 0.14 & 0.40 & 0.17 & 0.51 & 0.30 & 0.22 & 0.26 \\
 & A & \textbf{0.24} & 0.19 & 0.11 & 0.37 & 0.30 & 0.26 & 0.37 & 0.26 & 0.05 \\
 & PA (wrong labels) & \textbf{0.10} & 0.16 & 0.08 & $-$0.10 & 0.07 & 0.03 & 0.27 & 0.07 & 0.19 \\
\bottomrule
\end{tabular}
\end{table}

\paragraph{Forced-choice decoding and model scale.}
Without reasoning, the Qwen control showed the same pattern (\cref{tab:forced}). Few-shot agreement with wrong anchor labels matched agreement with correct ones (PF 0.24 versus F 0.23, $\Delta=-0.01$, $p=0.64$), and the anchored comparison did not beat few-shot scoring (A 0.24, $\Delta=+0.01$, $p=0.47$). Its stated relations were again less accurate than comparing its absolute score with the anchor's label (48\% versus 55\%). The missing calibration benefit is therefore not caused by the reasoning mode. With the larger DeepSeek-V4.1-Flash, neither registered test passed. Correct labels did no better than corrupted ones (F 0.26 versus PF 0.26, $\Delta=+0.00$, $p=0.45$, H7), and the anchored comparison did no better than few-shot scoring (A 0.24, $\Delta=-0.02$, $p=0.67$, H8). Its stated relations were also less accurate than comparing its absolute score with the anchor's label (45\% versus 55\%). DeepSeek reacted to the labels less than Qwen did. Corrupting them changed 10\% of its predictions, and the changed predictions followed the shifted labels no more often than chance (48\%, $p=0.64$). The absence of a calibration benefit thus holds for a second, larger model as well.

\paragraph{Errors of direct scoring (post hoc).}
To see where direct scoring fails, a separate analysis that was not registered examined its 320 test (child, item) scores. D was wrong on 175 (55\%). It over-scored far more often than it under-scored (112 versus 63) and placed most children at the middle level. For stereotyped language (A4) it gave 31 of 40 children a 1, although clinicians had scored 31 of them 0, and for social response (B9) it gave 37 of 40 a 1. We also asked the model, after scoring, to cite the transcript lines its score relied on, and checked these against the diarisation. Lines it attributed to the child often came from stretches dominated by the examiner. This flag, however, was about as common behind wrong scores as behind correct ones (46\% and 47\%). The model's one-sentence reason described a different level from its score in only 2\% of cases. The check depends on imperfect diarisation and on reasons given after scoring, which show what the model offers as support and not necessarily what it attended to. It found no link between flagged speaker attribution and scoring errors, but it cannot exclude that real attribution errors affect scoring. Agreement between the stated reason and the score also shows only internal consistency, not correct use of the clinical criteria. The clearest pattern in the errors is a compressed scale that separates children poorly.

\paragraph{Secondary analyses.}
Results on all 48 children were similar (\cref{tab:secondary}). For every arm, agreement was lower among the 11 test children of the first recording batch than among the 29 of the second, although the first-batch estimate rests on few children. Summed over the eight items, predicted and clinician scores correlated moderately for the main arms (\cref{tab:secondary}).

\section{Follow-up Analyses}
\label{sec:followup}
The main study found that clinician-scored examples did not improve agreement with clinicians and that, among the registered arms, which were not calibrated, direct scoring of the full transcript did best. Since added reference cases and added pipeline steps did not help, we asked where scoring hour-long sessions goes wrong. Three follow-up analyses address this. None of them was part of the main registration, and all were designed after its results were known. They differ in evidence level, which each subsection states.

\begin{figure}[t]
\centering
\includegraphics[width=\textwidth]{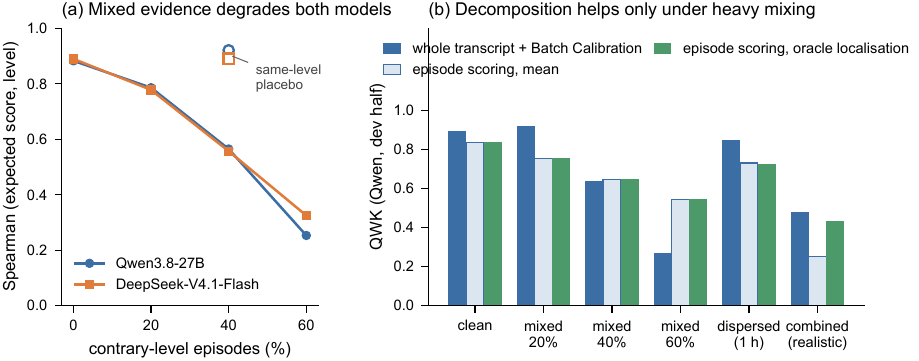}
\caption{\textbf{Follow-up analyses on synthetic sessions (\cref{sec:stress,sec:episodes}).} (a) Adding episodes of a contrary level to a vignette lowers both models' ranking of vignettes by level, more so at higher doses. Inserting the same amount of same-level text (open markers) has no such effect. (b) Development half of the synthetic sets, 48 sessions per condition. Scoring episodes separately and averaging them (light bars) beats calibrated whole-transcript scoring (blue) only under heavy mixing, and falls below it in the dispersed and combined conditions. Green bars keep only the episodes that the generator's records show to contain vignette text. This oracle is specific to the tested episode pipeline and is not a ceiling for every possible way of decomposing a session.}
\label{fig:followup}
\end{figure}

\subsection{Probability Read-out and Ordinal Calibration}
\label{sec:readout}

\paragraph{Question and evidence level.} Does the full-transcript model rank children better than its final scores suggest? This analysis reuses the 40 test children, so it is an additional evaluation of a cohort we had already analysed and not a confirmation. Its calibration baselines were added after we had seen the Qwen probabilities.

\paragraph{Design.} We re-ran D's prompt with forced-choice decoding, which gives a probability for each level, and compared the expected score under these probabilities with the most probable level and with existing read-out methods.

\paragraph{Finding.} The expected score ranked the test children well within each item (mean Spearman 0.54, permutation $p=0.0001$), compared with 0.36 for D's generated scores. Among pairs of children whose clinician scores differ by two levels, it ordered 91\% correctly. The most probable level, by contrast, piled up at the middle. For B9 it was level~1 for 95\% of children, even though the expected score still ranked them at 0.57. On clear vignettes written by us in D's format, the same prompt reached QWK 0.90 with no pile-up at the middle level, so the model applied the rubric correctly when the evidence was unambiguous. Existing label-free calibration applied to the probabilities raised mean QWK from D's 0.30 to 0.46 with Batch Calibration~\cite{zhou2024batch} and 0.45 with prior matching~\cite{liusie2023wordbias}. Per-item cut-points on the expected score, fitted leave-one-out on the clinician labels of the other 47 children, reached 0.50. The largest gains were on the items where predictions had piled up at the middle (B9 0.08 to 0.53, A8 0.21 to 0.66).

\paragraph{Limitation.} Batch Calibration and prior matching are transductive. They use no clinician labels, but they adjust each child's read-out with the model's outputs for the whole batch of test children, so no child is scored in isolation. The 0.50 result uses labels and is a different condition. With DeepSeek-V4.1-Flash, the expected score again ranked children better than the argmax (0.46 versus 0.38), but the 47-child calibration (0.38) was not significantly better than D ($\Delta=+0.08$, lower 95\% bound $-0.06$). The calibration gain is therefore shown for one of two models, on a cohort that was not independent.

\subsection{Controlled Stress Tests of Mixed Behavioural Evidence}
\label{sec:stress}

\paragraph{Question and evidence level.} Which property of an hour-long session weakens the model's graded judgement? This analysis uses synthetic sessions only, with no child data, and was pre-registered (design, metrics and decision rules) before its first model call.

\paragraph{Design.} We constructed 96 base synthetic vignettes: for each of the eight items, four minimal-pair sets of three short vignettes at levels 0, 1 and 2. The vignettes and the routine examiner-led filler were drafted with an LLM assistant from the rubric's level descriptors. No real transcript text was copied, and only aggregate statistics of the real transcripts were used to check realism. Each base vignette was rendered under 33 conditions, each adding one factor at a controlled dose. The factors were episodes at a contrary level (mixed evidence), fewer of the child's own episodes (sparsity), embedding in up to an hour of filler (dispersion), character-level recognition noise, windows cut regardless of utterance boundaries, interleaved examiner talk, and all of these together. For every factor there is a placebo condition that adds the same amount of text without the factor, for example episodes of the vignette's own level. Qwen3.8-27B and DeepSeek-V4.1-Flash scored every rendering by forced choice. The unit of analysis is the base vignette. The primary metric is the Spearman correlation between expected score and design level over the 96 vignettes in each condition. A factor counted as established only if its placebo-adjusted drop in this correlation had Holm-adjusted $p<0.05$ and a positive lower 95\% bound.

\paragraph{Finding.} Mixed evidence had the largest effect in both models, with a placebo-adjusted drop of $+0.36$ (95\% CI $+0.22$ to $+0.49$) for Qwen and $+0.34$ ($+0.20$ to $+0.47$) for DeepSeek, growing with dose (\cref{fig:followup}a). Sparsity was also established in both models ($+0.07$ and $+0.13$), and dispersion in DeepSeek only ($+0.22$). Noise, window cuts and interleaving were not established in either model. Both models gave a contrary minority more weight than its share. For items whose rubric counts how often a behaviour occurs, one atypical episode out of four moved the expected score 42--45\% of the way towards the all-atypical value. For A7, whose rubric credits the best example, the score stayed near the best example, which is what the rubric asks for. With all factors combined, the correlation fell to 0.22 (Qwen) and 0.11 (DeepSeek), lower than on the real transcripts (0.54 and 0.46), so the synthetic conditions cover the real one.

\paragraph{Limitation.} By construction, the design level of a mixed vignette is its majority level. Real ADOS-2 items are not all scored by majority. Some weigh specific behaviours or their frequency, and some the quality of the whole interaction, so part of the measured drop may be clinically correct. The finding is that mixing behaviour of different levels caused a large, dose-dependent loss of ranking in this synthetic task, in both models. It does not show that real ADOS-2 scoring fails mainly because of mixed evidence. That would need clinician-coded mixed cases.

\subsection{Why Episode-Level Decomposition Falls Short}
\label{sec:episodes}

\paragraph{Question and evidence level.} If the model over-weights a contrary minority, does scoring each episode on its own and combining the scores with an explicit rule help? This analysis uses the synthetic sessions of \cref{sec:stress}. The method and its decision rules were pre-registered. The analysis of why it failed is exploratory.

\paragraph{Design.} We scored each episode separately and combined the episode scores with an item-specific rule: a mean for items that count how often a behaviour occurs, a trimmed mean for items about typical behaviour, and the best example for A7. Episodes are contiguous runs of one protocol activity, split into chunks of at most 12 windows. A futility rule on the development half of the synthetic sets decided whether the held-out half would be run. The method had to beat calibrated whole-transcript scoring on both the mixed-evidence conditions and the combined realistic condition. A second pre-registered variant would first remove episodes that a model probe judged irrelevant.

\paragraph{Finding.} Under heavy mixing the method did better. At 60\% contrary episodes it reached QWK 0.59, against 0.27--0.32 for whole-transcript scoring with Batch Calibration or prior matching, and over the three mixing doses 0.71 against 0.66. In the combined condition none of the eight configurations came close (best 0.31 versus 0.48--0.52), so the study stopped before its held-out half. The second variant stopped at its localisation gate, because the probe separated evidence-bearing from filler two-window units with an AUROC of only 0.74, below the required 0.80. The generator keeps a record of which transcript windows carry vignette text, which shows where the loss occurs (exploratory). Episodes that contain vignette text still separated level~2 from level~0 within each item, with AUROC 0.78 in the combined condition and 0.90 with dispersion alone, while filler-only episodes did not (0.51). The useful episodes, however, are rare and mostly filler. In the combined condition only 1.4 of 19 episodes per session contained any vignette text, and in those the vignette made up a median 11\% of the text. Averaging only those episodes, which no practical localiser can do, raised QWK from 0.25 to 0.43, still below calibrated whole-transcript scoring (0.48). With dispersion alone this oracle gained nothing (0.73 versus 0.85, \cref{fig:followup}b). Pooling with an upper quantile or the maximum instead of the mean did worse than the mean in both long conditions.

\paragraph{Limitation.} These results apply to the episode pipeline we tested, at its granularity, on synthetic sessions. Within that pipeline, splitting a session spread sparse evidence thinner instead of isolating it, and even perfect localisation did not reach the whole-transcript baseline. Other ways of decomposing a session may behave differently.

\section{Discussion and Limitations}
\paragraph{Sensitivity and benefit.}
Whether a model is \emph{sensitive} to the examples' labels and whether correct labels bring a \emph{calibration benefit} are separate questions, and here they have different answers. The model did react to the labels. Corrupting them changed about a third of the few-shot predictions, more often towards the corrupted labels than chance would give. Yet with correct labels agreement was not higher than with shifted labels or with no examples at all. This is an absence of a detectable improvement, not proof that the effect is zero, but any benefit large enough to matter for 40 children was not seen. The labels moved individual predictions, but not, on balance, towards the clinicians. Qwen without reasoning and the larger DeepSeek model behaved the same way, and DeepSeek reacted to the labels even less. Systems that show an LLM scored examples to align it with experts should therefore include a corrupted-label control before claiming a benefit, since a model can react to labels without adopting the experts' scale.

\paragraph{Why comparison did not help.}
We built the anchored comparison on the premise, taken from our video pilots, that relative judgements would be more reliable than absolute ones. For transcripts the mechanism test does not support it. Relations the model stated between two children were right no more often than relations implied by its absolute scores, and less often than its absolute score compared with the anchor's true label. On this data the model placed a child against the rubric better than against another child's evidence. We did not test why. One possibility is that score-free evidence summaries of different children read alike. Another is that judging two sets of evidence at once is harder than judging one against written criteria. Both remain hypotheses.

\paragraph{Where the signal is.}
Reciprocal communication (B10) was among the best-rated items in every arm. Conversation (A8) did well in the evidence-based arms, and rapport (B11) and restricted interests (D4) under direct scoring. Summed scores over the eight items correlated moderately with the clinicians' sums. Agreement was much lower in the first recording batch, which has few test children. Overall agreement is modest, and without a second clinician rating the same sessions we cannot say how far it falls short of agreement between human raters.

\paragraph{Less structure, more agreement.}
Among the registered designs, none of which was calibrated, direct scoring of the full transcript was the most accurate. In exploratory comparisons its advantage over scoring of extracted evidence came from the five items defined over the whole session. For those items a handful of evidence snippets is a poor summary of an hour of interaction, and the long-context model did better reading everything. For items tied to specific activities, extraction cost nothing. Pipelines that condense long sessions before judging them should be compared with direct reading, item by item. The follow-up analyses (\cref{sec:followup}) agree. On synthetic sessions, scoring episodes separately and aggregating them fell below calibrated whole-transcript scoring whenever the evidence was sparse within an hour, even with oracle localisation. Among the Qwen read-outs we tested, the best combined whole-transcript scoring with existing batch calibration, although that result comes from a cohort that was not independent and did not replicate with DeepSeek. In the synthetic stress tests, mixing behaviour of different levels was the condition that hurt both models most. How that maps onto real ADOS-2 scoring, where items are not all scored by majority, remains open. New remedies should be tested against calibrated whole-transcript scoring, not against uncalibrated direct scoring.

\paragraph{Conclusion.}
In this study, expert-scored examples did not calibrate an LLM rater of ADOS-2 transcripts, and more elaborate evidence processing lost session-level information. Claims that examples calibrate such raters, and claims that added pipeline steps help, should be checked against corrupted-label controls and calibrated whole-transcript scoring.

\paragraph{Scope and limitations.}
The study evaluates item coding of eight speech-based ADOS-2 items, not autism diagnosis, and it did not evaluate clinical usefulness. No clinician used the model's outputs, and none of the agreement levels reported here has been validated on an independent cohort. The study uses one dataset, 48 children assessed with ADOS-2 Module~3 in Chinese and recorded in two batches with different score distributions. Each child has one set of clinician scores, so we cannot estimate agreement between human raters on this sample, which limits what any automatic rater can reach. We study only the eight items that can be judged from speech. The camera position rules out the visual items. The transcripts carry no speaker labels, so the model has to infer which utterances are the child's and inherits the recogniser's errors. The method needs a pool of clinician-scored children assessed with the same protocol to draw anchors from. We evaluate one anchor per score level, drawn by a fixed rule, and do not study how agreement depends on the choice or number of anchors. The main arms use one LLM, and the forced-choice replication adds one more model. The pilot evidence in \cref{sec:challenge} rests on model judges and pre-registered tests, not on clinician review of the flagged segments. The calibration results in \cref{sec:followup} reuse the 40 test children and replicate for only one of the two models, so they need an independent cohort. Finally, the synthetic sessions were drafted with an LLM assistant and are cleaner than real behaviour, and their design levels define the correct answer for a mixed vignette by majority, which a clinician might not accept for items that credit occasional behaviour.

\paragraph{Data and privacy.} The recordings and transcripts are clinical data of minors and are not released. All model scoring of child data ran on servers controlled by the research team, without third-party APIs. The synthetic sessions of \cref{sec:followup} contain no child data. No transcript excerpt is reproduced in this paper.

\paragraph{Ethics.} The study was reviewed as human-subjects research by the Ethics Subcommittee of the Scientific Research Ethics Committee, School of Politics and Public Administration, South China Normal University, which found its content and protocol consistent with research-ethics requirements and approved its continuation on 12 November 2026 (approval no.~SCNU-PAP-2026-006).

{\small
\bibliographystyle{plainnat}
\bibliography{references}
}

\clearpage
\appendix
\section{Pilot Experiments in Detail}
\label{app:pilot}

\paragraph{Predicting totals from video.}
Segment features from a video encoder (VideoMAE~V2~\cite{wang2023videomaev2}), turned into per-child anomaly scores and aggregated by ridge regression, failed all four pre-registered tests of predicting the ADOS total (best permutation $z=+1.06$, $p\ge0.167$). Always predicting the training-set mean had a lower mean absolute error (4.73) than every video-based method tried (4.96--5.63). Absolute 0--4 behaviour ratings from a vision-language model for each 16-second segment, aggregated per child, were no better than the mean (MAE 4.83--4.85).

\paragraph{Signal within a child.}
In a blind test, a vision-language model saw four segments the system ranked as most atypical and four random segments from the same child, in shuffled order, and rated each for atypical social behaviour on 0--10. Flagged segments scored 0.64 points higher (Wilcoxon $p=0.0106$, $n=45$ children), and the effect persisted after removing the first two or five minutes of each session. Replacing the judge with an unrelated, much smaller model (Gemma-4-E4B) preserved the effect ($+0.50$, $p=0.0044$, $n=46$).

\paragraph{Comparison across children.}
Together these results suggest that segment-level model judgements order segments within a child but do not place children on a common scale. A diagnostic points the same way. Learned segment features identified the source recording with 79--85\% accuracy, so cross-child comparisons mostly reflect who and where, and less the behaviour.

\paragraph{Adding features.}
Three speaker-agnostic transcript statistics predicted the total (MAE 4.04, $z=+3.03$, $p=0.004$ after correction for nine feature families). Adding vision, pose or speaker-separated interaction features to them weakened or removed the signal, and speaker-separated interaction features failed all twelve pre-registered item-level tests. With $n=48$, learning a richer mapping from features to scores did not succeed in these pilots.

\end{document}